\documentclass[doublecol]{epl2} 
\usepackage{graphicx,epsf,subfigure}
\usepackage{latexsym,amssymb}
\usepackage{float}
\title{Tsunami generated by a granular collapse down a rough inclined plane}
\shorttitle{Tsunami generated by a granular collapse down a rough inclined plane} 

\author{S. Viroulet\inst{1} \and A. Sauret\inst{2} \and O. Kimmoun\inst{1}}
\shortauthor{S. Viroulet \etal}

\institute{                    
  \inst{1} IRPHE, CNRS UMR 7342 - Aix-Marseille University - AMU - Ecole Centrale Marseille, France\\
  \inst{2} Department of Mechanical and Aerospace Engineering, Princeton University, Princeton, New Jersey 08544, USA
}
\pacs{47.57.Gc}{Granular flow, complex fluids}
\pacs{92.10.hl}{Tsunamis}
\pacs{45.70.Ht}{Avalanches (granular systems)}

\abstract{In this Letter, we experimentally investigate the collapse of initially dry granular media into water and the subsequent impulse waves. 
We systematically characterize the influence of the slope angle and the granular material on the initial amplitude of the generated leading wave and the evolution of its 
amplitude during the propagation. The experiments show that whereas the evolution of the leading wave during the propagation is well predicted by a solution of the linearized Korteweg-de 
Vries equation, the generation of the wave is more complicated to describe. Our results suggest that the internal properties of the granular media and the interplay with 
the surrounding fluid are important parameters for the generation of waves at low velocity impacts. Moreover, the amplitude of the leading wave reaches a maximum value at 
large slope angle. The runout distance of the collapse is also shown to be smaller in the presence of water than under totally dry conditions. This study provides a first insight 
into tsunamis generated by subaerial landslides at low Froude number.}

\begin{document}

\maketitle

\begin{center}
\begin{figure*}
\begin{center}\includegraphics[width=18cm]{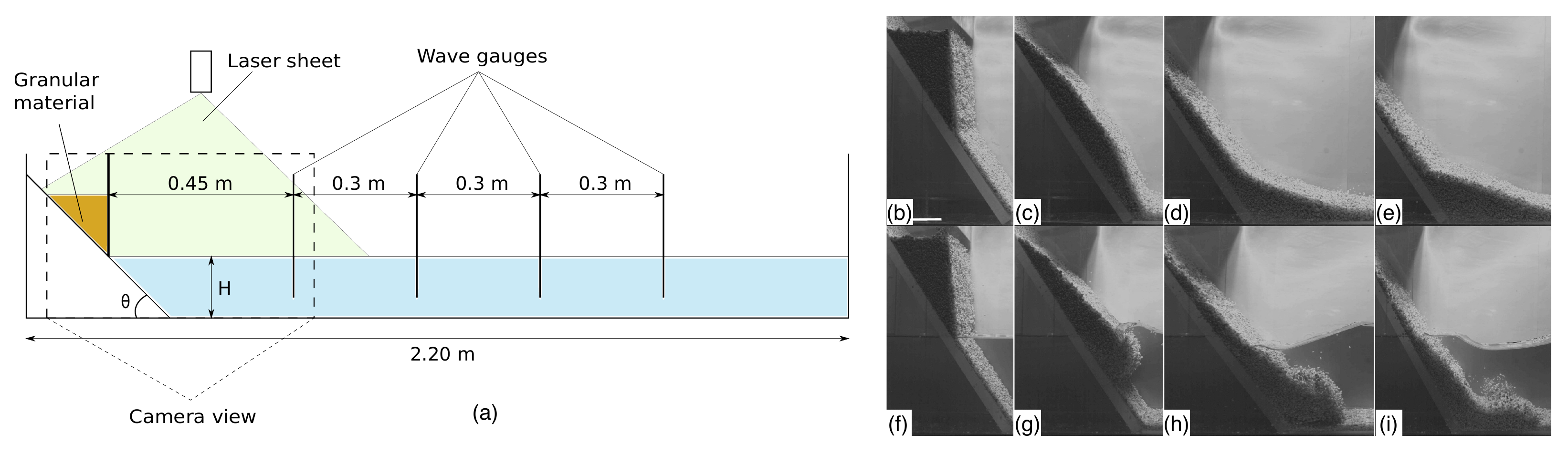}\end{center}
\caption{Schematic view of the experimental setup. The shape of the granular flow and the generation of the waves are recorded by the high speed camera. 
The evolution of the amplitude is measured with four contact-type wave gauges. Successive instants of the collapse of the granular material ($m=2$ kg) (b-e) in the dry case, and (f-i) in presence of water. 
The pictures are taken every $0.2$ seconds and the scale bar is 5 cm.}
\label{fig:1}
\end{figure*}
\end{center}
In the last decade, two particularly destructive tsunamis, in the Indian Ocean \cite{titov2005} and in {Pacific Ocean} \cite{mori2011}, have raised the necessity to predict, or at least evaluate the 
hazards induced by such events for the population and human activities. Although many tsunamis arise from underwater earthquakes \cite{grilli2012}, some are also induced by submarine and subaerial 
landslides which can be locally more dangerous as they form near the coast and may exhibit extreme runup \cite{kremer2012}. 

One approach  to study tsunamis generated by subaerial landslides relies on the simple configuration of a solid block impacting the water \cite{walder2003,enet2007}. 
During a landslide, strong interactions occur between the slide and the generated waves. In this case, estimating real situations requires a deformable slide \cite{fahad2012}. 
Most previous studies focused on the particular case of a Froude number $Fr=v_s/\sqrt{g\,H}$ larger than one, i.e. on a slide velocity ${v_s}$ larger 
than the shallow water wave velocity $\sqrt{gH}$ ($g$ being the gravity, $H$ the water depth). This configuration is relevant to model geophysical events such as the 
mega-tsunami of Lituya Bay, Alaska (1958)  where the estimated Froude number was about $Fr\simeq 3.2$ \cite{fritz_alaska}. However, in the simple situation of a cliff 
collapse, which is expected to happen in the South of France  \cite{recorbet2010}, the dynamic would be different as the Froude number becomes smaller than unity. In this 
situation, an estimate of the interplay between the granular material and the wave is required. Despite important implications on risk control, landslides for a Froude number 
smaller than one and the generation of the subsequent tsunamis have been poorly investigated.

The lack of such studies may be explained by the difficulty of modeling dense flows of granular media despite their major stake in industrial and geophysical applications 
\cite{iverson1997,iverson2010}. Due to these applications, a huge number of studies considered various dense flows of dry granular materials \cite{pouliquen1999,gray2001,balmforth,annrev}. 
The mechanical properties of the grains are not completely understood but recently, Jop et al. \cite{jop2006} suggested a continuum model, the $\mu\textrm{(I)-rheology}$, to account for the unique 
properties of dry dense granular flows. This rheology gives a good description in many situations such as a granular collapse \cite{lacaze2009,lagree2011} or granular flows on a pile 
\cite{jop2006}. However, the addition of water drastically changes the dynamics of granular flows as emphasized by recent studies with partially wet granular media 
\cite{nowak2005,huang2005,herminghaus2005,fiscina2010,chopin2011}, suspension of particles \cite{boyer,ancey1,ancey2} and underwater granular  flows \cite{cassar2005,rondon2011}. These studies show 
the influence of both the capillary effects and the surrounding fluid, but they also emphasized the difficulty in predicting the dynamics of the granular media and their interplay with the fluid. In particular, 
underwater granular events are complicated to describe due to the contradictory effects of the drag force and the lubrication of the grains \cite{topin2012}. It becomes even more complex to describe the 
behavior of the impact of dry granular media from the air into water and the subsequent generated waves.

In this Letter, we rely on extensive experiments to analyze the interactions between the granular flow and the  waves generated by the landslide at Froude numbers smaller than 
one. Typically, in our experiments $Fr \sim 0.2 -0.8$. The present situation is relevant to model tsunamis generated by cliff failures located just above the sea surface, 
which are characterized by low impact velocities.



The experimental setup is sketched in Fig. \ref{fig:1}(a). It consists of a glass tank of dimensions $2.20\ \textrm{m}$ long, $0.40\ \textrm{m}$ high and $0.195\ \textrm{m}$ 
wide. The water depth considered in this study is $H=15 \,\textrm{cm}$. Three different granular materials are used: spherical glass beads of diameters $4$ and 
$10\ \textrm{mm}$ and density $\rho_{GB}\simeq 2500\ \textrm{kg.m}^{-3}$, and a non-spherical aquarium sand of mean diameter $4\,\pm\, 1\ \textrm{mm}$ and density 
$\rho_{AS}\simeq 2300\ \textrm{kg.m}^{-3}$. The slope is made of a polyvinyl plate roughened by gluing one layer of the considered granular media onto it to ensure no-slip boundary conditions at the bottom.  A vertical gate, opened vertically, 
delimits the initial reservoir of granular materials located just above the undisturbed free surface. The generation of the wave and the evolution of the granular flow are recorded using a high speed
camera at $100\ \textrm{fps}$ (Phantom V641, 4M Pixels). The shape of the granular material is highlighted using a laser sheet located above the free surface. The evolution of the free surface elevation is 
measured using four resistive wave gauges with a sampling frequency of $200\ \textrm{Hz}$, which is sufficient to capture the dynamics of the waves. The gauges are located at $0.45\ \textrm{m}$, $0.75\ \textrm{m}$, 
$1.05\ \textrm{m}$ and $1.35\ \textrm{m}$ from the lifting gate.

Typical experiments are illustrated by successive snapshots of the granular collapse in a dry case (Fig. \ref{fig:1}(b)-(e)) and in presence of water 
(Fig. \ref{fig:1}(f)-(i)) for an initial mass $m=2$ kg of aquarium sand and a slope {$\theta=50\!\char23$}. It shows that the presence of water modifies the shape and the 
timescale of the granular front during the collapse, which occurs in an inertial regime \cite{gondret2003,cassar2005,rondon2011}. Also note that the time scale of 
the granular collapse in air is smaller than in the presence of water and that their morphologies are different (see Fig. \ref{fig:1}(c) and \ref{fig:1}(g) taken at the 
same instant). In water, the surrounding fluid generates an important drag on the granular material which leads to the generation of a larger front. At the same time, a wave 
is generated at the surface on a timescale similar to the granular collapse, and thus it is not possible to decouple the two different phenomena.


\begin{center}
\begin{figure}
\begin{center}\includegraphics[width=8cm]{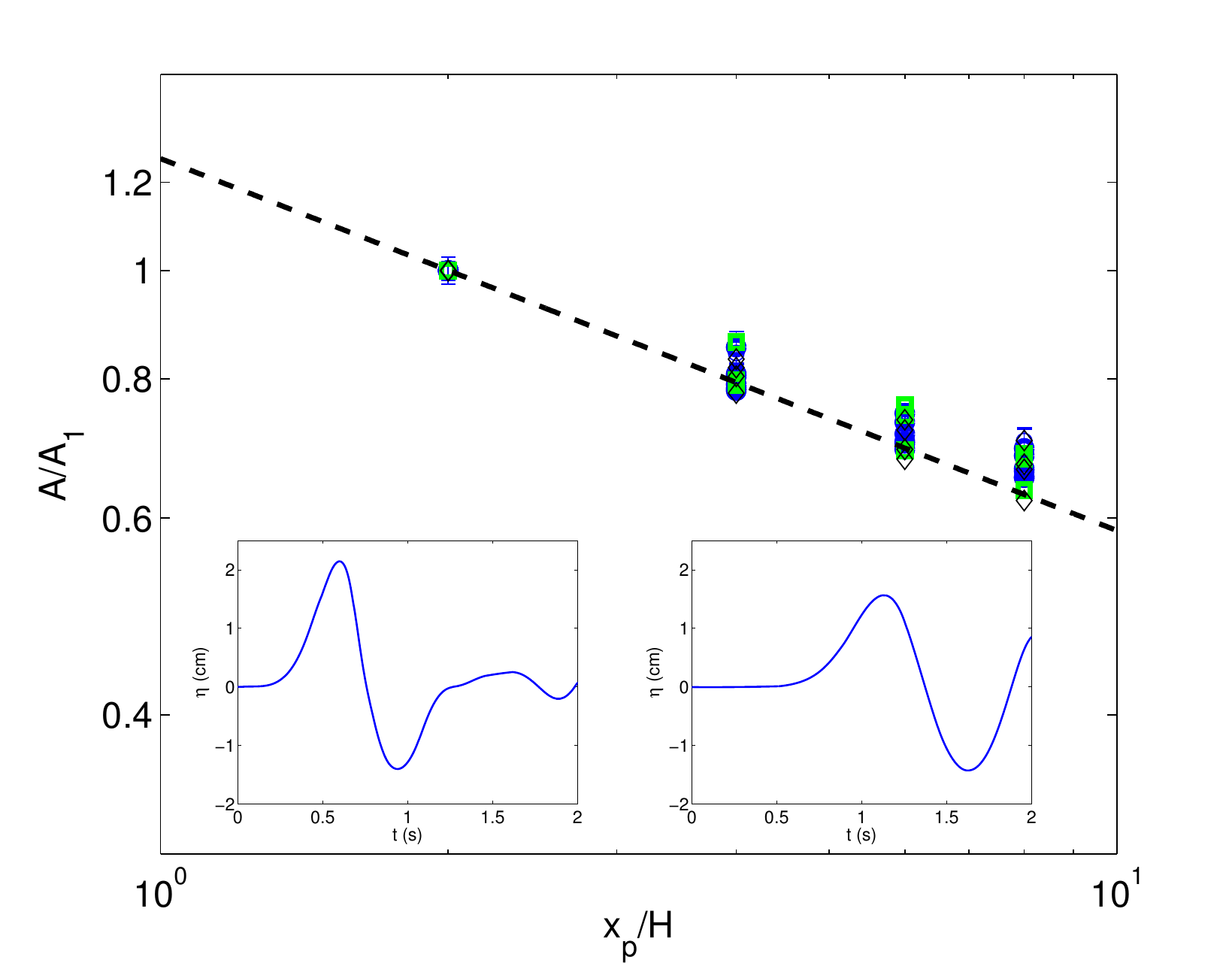}\end{center}
\caption{Evolution of the dimensionless amplitude of the generated leading wave in logarithmic scale for various masses and various slopes using glass beads of $4$ mm 
($\circ$), $10$ mm ($\square$) and aquarium sand ($\diamond$). The dashed line is the scaling law $x^{-1/3}$. 
The uncertainties are plotted on the blue circles. Inset: evolution of the wave amplitude $\eta$ at the first and third probes for $m=2$ kg of $4$ mm glass beads and a slope of $45\!\char23$.}
\label{fig:2}
\end{figure}
\end{center}

Let us consider the propagation of the leading wave, which is also the largest one, and is followed by an oscillatory wave train. We assume that the initial 
amplitude of the leading wave is known and we focus here on its propagation. We shall see in the following how the slope influences the amplitude of the leading wave. 
We ensured that the experiments are completely reproducible. We report  in figure \ref{fig:2} the evolution of the maximum amplitude of the generated leading wave as 
a function of the distance to the collapse for various slope angles, initial mass and different granular materials. For each experiment, we rescale the amplitude at the 
probes by the amplitude $A_1$ at the first probe. To explain the dependence of the tsunami amplitude with the distance from the source, $x$, we consider the weakly 
dispersive linear shallow water approximation. In this approximation, the free surface elevation $\eta$ is described by the the linearized Korteweg-de-Vries equation 
\cite{whitam}:
\begin{equation}
 \frac{\partial{\eta}}{\partial{t}}+c\,\frac{\partial{\eta}}{\partial{x}}+\frac{c\,H^2}{6}\frac{\partial{^3\eta}}{\partial{x}^3}=0.
\end{equation}
A solution to this equation is
\begin{eqnarray}\label{sol_Airy}
\eta(x,t)=Q\left(\frac{2}{c\,t\,H^{2}}\right)^{1/3} \textrm{Ai} \left[\left(\frac{2}{c\,t\,H^{2}}\right)^{1/3}(x-c\,t)\right],
\end{eqnarray}
where $Q$ is a constant corresponding to the cross-section of the water volume displacement, $H$ is the water depth, $c=\sqrt{gH}$ and $\textrm{Ai}$ is the Airy function. Thus, during the propagation, 
the amplitude of the leading wave decays as $t^{-1/3}$, which leads to a dependence as $x^{-1/3}$. This scaling law agrees with the experimental data regardless of the initial mass or slope 
(see Fig. \ref{fig:2}). Note also that the viscosity and the interfacial tension have negligible effects on the propagation of the wave as observed for solid body impact 
\cite{viroulet2013} or granular impact at $Fr>1$ \cite{fahad2012,heller2008}. Thus, using a far-field approximation, we can estimate the amplitude of the waves during its 
propagation.

\begin{center}
\begin{figure}
\begin{center}\includegraphics[width=8.5cm]{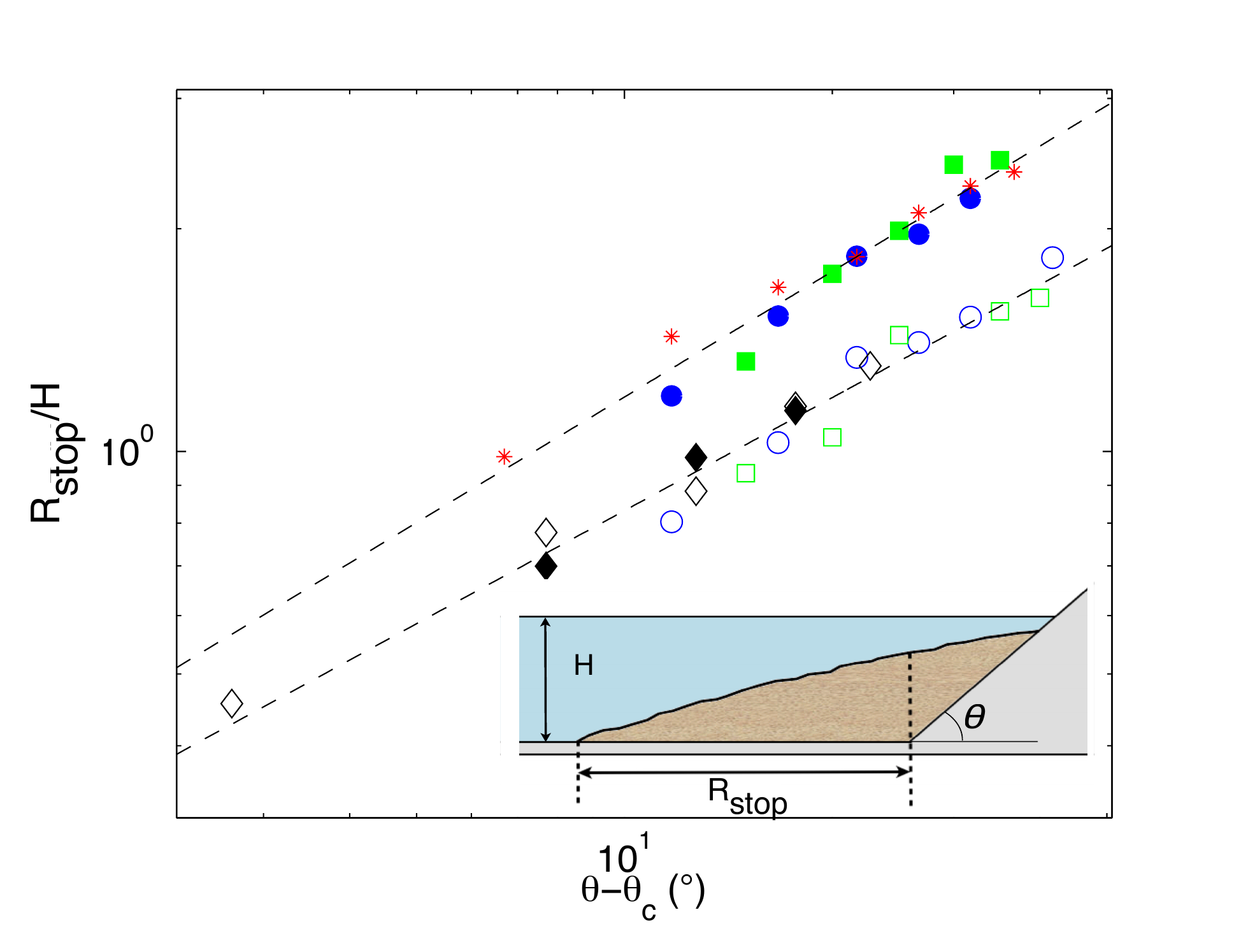}\end{center}
\caption{Dimensionless runout distance as a function of the slope angle for the three different granular media: glass beads $10$ mm ($\square$), $4$ mm ($\circ$) and aquarium sand ($\diamond$). 
Full symbols are the results for a dry granular collapse and empty symbols are in presence of water. The red asteriks are underwater collapse using $4$ mm glass beads. The two dashed line are the best 
fits for the two regimes. Inset: schematic of the situation.}
\label{fig:3}
\end{figure}
\end{center}

 We now need to estimate the influence of the different parameters, such as the slope angle or the mass of granular media, on the initial amplitude of the wave. In the 
 following, we focus on the influence of the granular flow properties on the wave generation. The evolution of the run-out distance of the granular flow {with the slope angle 
 $\theta$} for both dry collapse and in presence of water is shown on Fig. \ref{fig:3}. In our experiments, the aspect ratio of the slide is related to the slope angle 
 by ${h}/{L}=\tan\theta$, where $h$ is the initial height of the slide and $L$ is the initial length. Previous studies have considered the run-out distance on a slope \cite{staron2008,staron2009},
but only in a dry case. In our situation, we clearly see that the data is separated into two main curves.  
The run-out distance associated to the collapse of glass beads without water, as well as the underwater collapses, are all located on the upper curve. 
It suggests that for spherical materials, the runout remains the same if the collapse occurs in one phase only (air or water). Indeed, in our case the viscous 
effect is negligible compared to the inertia of the particles. However, we also notice that the runout distances associated with the collapse of glass beads from 
air into water collect in the lower curve. This is likely due to the presence of air captured during the collapse which decreases the velocity of the slide. 
Also, the interaction between the generated wave and the granular collapse can have a non-negligible contribution. Note that the aquarium sand has a peculiar 
behavior as the run-out distance is the same in all the situations, which may be due to the coarse shape of the aquarium sand that limits the velocity of the particles. 
The runout distance follows a power law $R_{stop}=\lambda (\theta-\theta_c)^{\alpha}$ with $\lambda=0.32$, $\alpha=0.56$ 
and $\lambda=0.26$, $\alpha=0.51$ for dry and into water collapse respectively. The critical angle $\theta_c$ is the angle of avalanche measured experimentally 
\cite{borzsonyi2008}. The value of the critical angle for each granular material is represented in table \ref{table_angle}. The contradictory effects of the ambient fluid observed by Topin et al. \cite{topin2012} for an underwater collapse which can enhance the granular flow 
are not observed in the configuration of dry impact into water in the inertial regime. 
Note that in our experimental setup, dry granular collapse and underwater collapse leads to a similar runout distance. But when the granular material impacts the free-surface 
of the water, the runout is smaller. {Indeed, a larger amount of energy is dissipated by internal friction
because of the presence of air captured in the granular material.} As observed by Topin et al. \cite{topin2012}, the added hydrodynamic forces of the ambient fluid on the 
granular material increase and so the duration of the collapse is larger for impact into water. Note also that the difference between wet and dry collapse is likely due to 
the speed at which the granular materials spread over the bottom. It also explains why the aquarium sand have a peculiar behavior in our experiments. Its runout is the same 
in both the dry case and in presence of water because the velocity of the granular front remains much lower for aquarium sand than for spherical glass beads. It seems to indicate 
that for the aquarium sand, the friction on the bottom and between the different layers prevails over the hydrodynamic forces of the ambient fluid.

{
\begin{table}[h!]
  \begin{center}
   \begin{tabular}{|c|c|c|c|}
   \hline
   granular material & diameter (mm) & $\theta_c$   \\
   \hline
   small glass beads & $1.5$  & $25,7^{\circ}\pm 0,9^{\circ}$ \\[1mm]
   \hline
   medium glass beads & $4$ & $23,3^{\circ}\pm 0,8^{\circ}$ \\[1mm]
   \hline
   large glass beads & $10$ & $20,1^{\circ}\pm 1,2^{\circ}$ \\[1mm]
   \hline
   aquarium sand & $\simeq 4$ & $37,3^{\circ}\pm 0,6^{\circ}$  \\[1mm]
   \hline
   \end{tabular}
  \caption{Mean particle diameter, $d$ and critical angle of avalanche $\theta_c$ for the four different granular media.\label{table_angle}}
 \end{center}
\end{table}}

\begin{center}
\begin{figure}
\includegraphics[width=8.8cm]{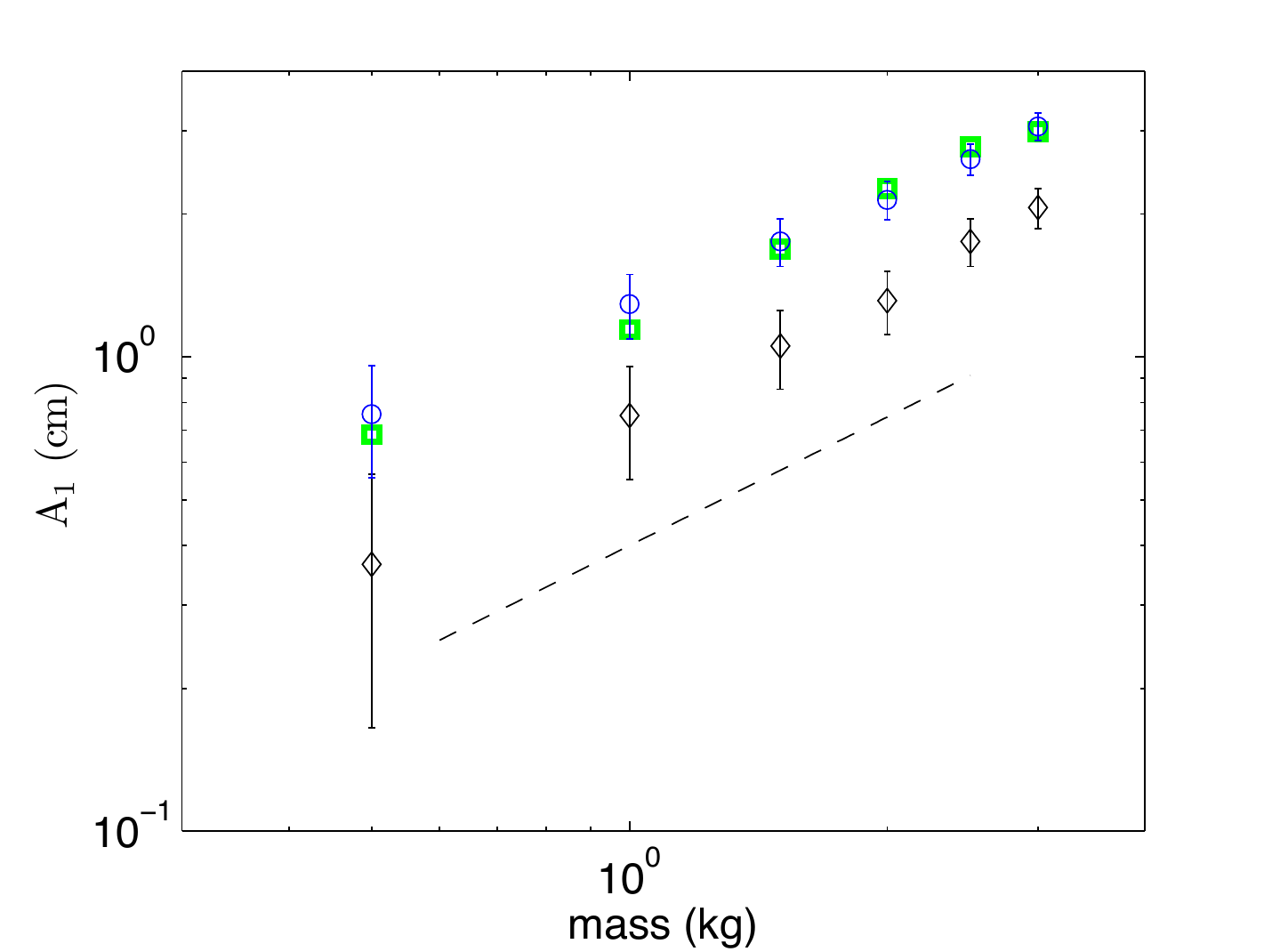}
\caption{Evolution of the amplitude at the first probe for a slope angle of $45^{\circ}$, as a function of the initial mass for the $10$ mm glass beads ($\square$), $4$ mm ($\circ$) and aquarium sand ($\diamond$). The error 
bars are illustrated on the data for $4$ mm glass beads and the aquarium sand. The dashed-line is the best fitting slope $m^{0.9}$.}
\label{fig:4}
\end{figure}
\end{center}

The amount of rocks collapsing into the water is a relevant parameter for geophysical events and is thus also considered experimentally. Fig. \ref{fig:4} reports the 
influence of the initial mass of granular material on the wave amplitude. The slope angle remains constant at $45^{\circ}$ and the three different granular materials are considered. 
From the previous results we can restrict our study to the first wave gauge since the later propagation of the wave can be estimated by the shallow-water model. In all of the 
experiments, the volume fraction of the granular media is estimated to be $0.60\pm0.05$. Note that this value is estimated to be slightly larger for the aquarium sand because 
of the non-sphericity of the particle and the more important polydispersity. However, the larger compaction of the aquarium sand associated with its lower density leads to 
an equivalent density of the granular slide $\rho_{eq}=\phi\rho_g$ for the three different granular materials, where $\phi$ is the volume fraction and $\rho_g$ is the density 
of the granular material. The amplitude of the wave at the first probe increases with the initial mass following a power law $A_1\propto m^{\alpha}$ with 
$\alpha\simeq 0.9\pm 0.05$ for the three granular media. Note that predicting the exact dependence of the wave amplitude on the mass of the collapse remains a challenging 
issue for waves generated at low impact velocity. Indeed, in previous studies \cite{fritz2004,fahad2012}, the Froude number at the impact was always larger than unity such 
that all the initial amount of granular material contributed to the generation of the leading wave. However, in the present study, when the leading wave starts propagating part of 
the granular material is still collapsing along the slope for large mass of the slide. Therefore, all the initial mass does not generate the first wave and future experiments 
will have to consider the time scale of the collapse compared to the time scale of the wave generation. Note also that in our configuration, we were only able to use glass 
beads of the same density. We are, thus, not able to tune the mass of the material without modifying the volume accordingly. Further experiments using beads of different density would be of 
interest to investigate the relative influence of the initial mass and the initial volume of the slide.

\begin{center}
\begin{figure}
\begin{center}
\includegraphics[width=8.8cm]{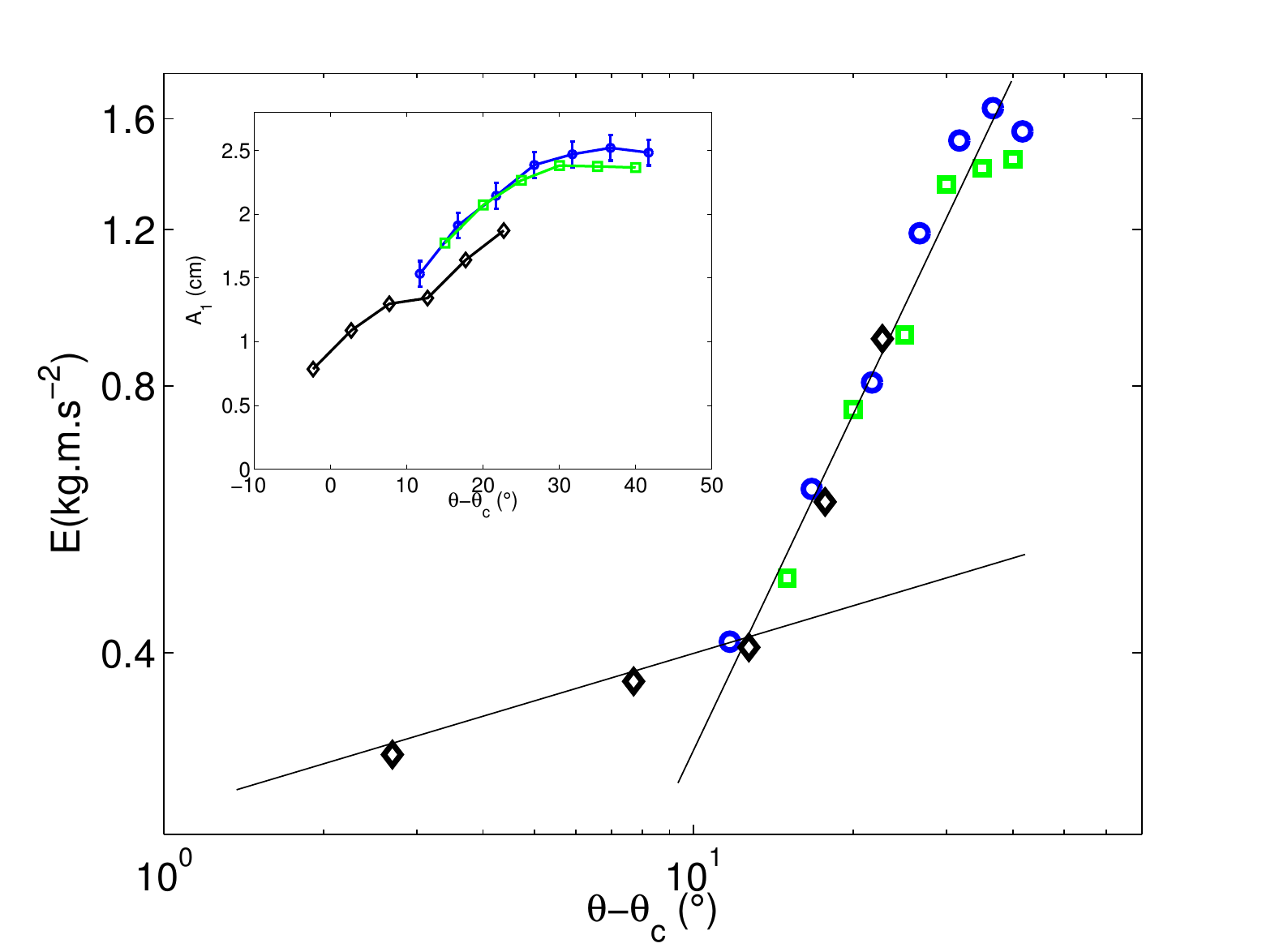}
\end{center}
\caption{Evolution of the energy of the generated wave train as a function of the slope angle for an initial mass of $2\ \textrm{kg}$. Inset: evolution of the amplitude at the first wave gauge as a 
function of the slope angle for an initial mass of $2\ \textrm{kg}$. In both figures ($\circ$) are for 4 mm glass beads, ($\square$) for 10 mm and 
($\diamond$) for the aquarium sand.}
\label{fig:5}
\end{figure}
\end{center}

\begin{center}
\begin{figure}
\begin{center}
\includegraphics[width=8.8cm]{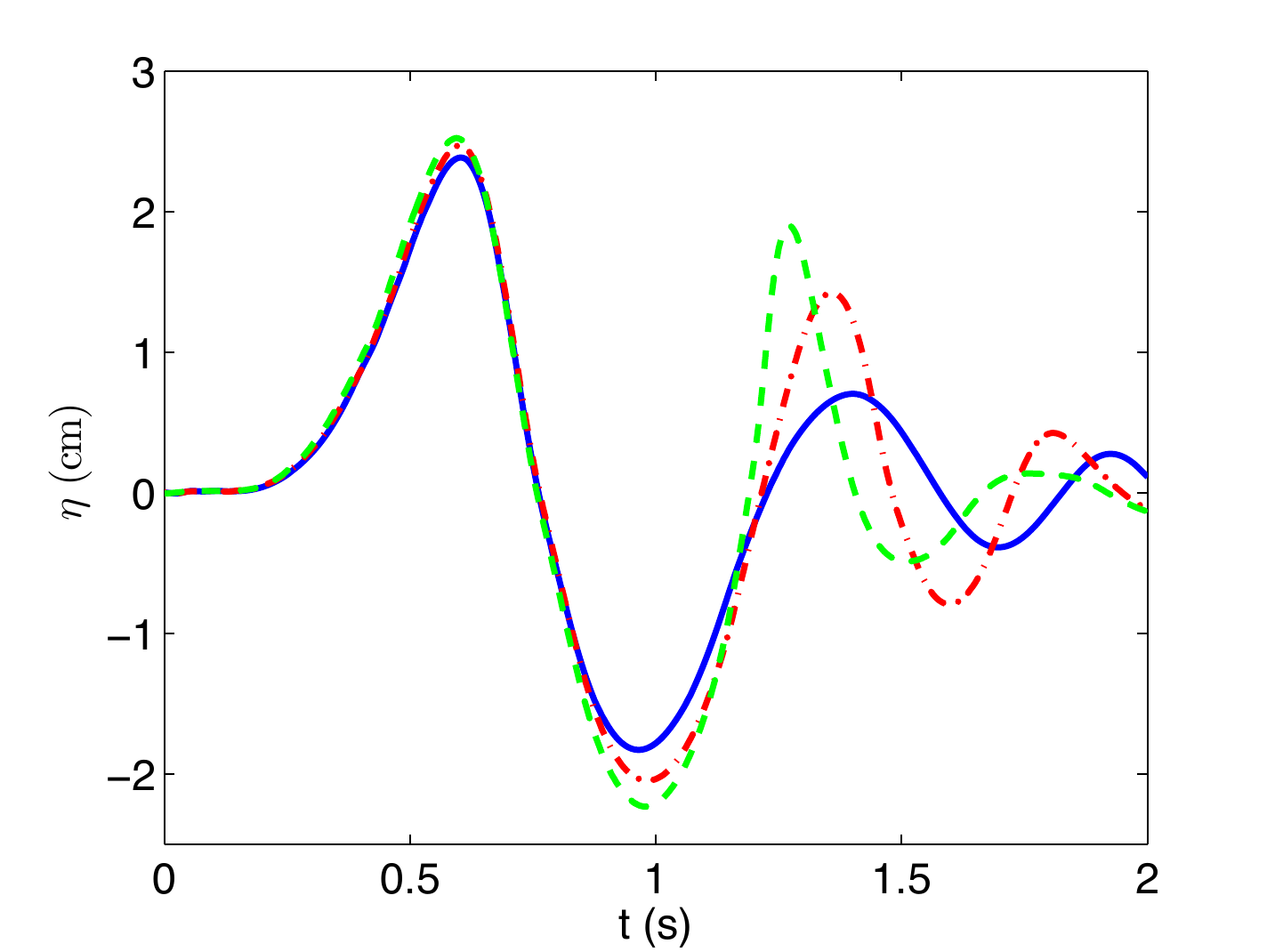}
\end{center}
\caption{Evolution of the free surface at the first probe as a function of the slope angle with the glass beads of $4\ \textrm{mm}$, at an angle of $50^{\circ}$ (continuous line), 
$55^{\circ}$ (dash-dotted line) and $60^{\circ}$ (dashed line) .}
\label{fig:6}
\end{figure}
\end{center}

In previous studies, the velocity of the slide was large enough to neglect the influence of the slope and the deformation of the granular media \cite{fritz2004,fahad2012}. Here, the velocity 
of the granular materials at the impact is close to zero and the interplay between the fluid and the beads is important. Therefore, the previously obtained results cannot be 
used to estimate the amplitude of the leading wave generated by cliff failure. We have reported in Fig. \ref{fig:5} the influence of the slope on the maximum amplitude at 
the first wave gauge for an initial mass of $2\ \textrm{kg}$. For a given granular material, the amplitude increases with the slope. 
In addition, our experimental results suggest that the relevant parameter to describe the influence of the slope is the difference between the slope and the critical angle of 
the considered granular media. Using this new variable, the results collapse quite well for the different granular media. It also shows a saturation at large slope, 
$\theta-\theta_c>30^{\circ}$, where the amplitude of the leading wave reaches a maximum. However, even if the amplitude of the wave saturates, increasing the slope still 
transfers energy to the waves as illustrated in figure \ref{fig:5}, where we have evaluated the evolution of the energy contained in the wave train for all the granular 
materials. The energy per unit width is estimated by the relation $E=2\,E_p=\rho g c\int\eta^2\mathrm{dt}$ where $\eta$ is the free-surface elevation. The results collapse onto 
a master curve and a transition is visible around $\theta-\theta_c \simeq 10 \!\char23$. Note that a given amplitude for the first wave does not 
necessary lead to the same wave train and thus to the same total energy associated with the tsunami. Indeed the behavior of the collapse is very different between 
each granular material for the same $\theta-\theta_c$ so the generated wave train will be different. $50\%$ to nearly $80\%$ of the energy of the 
wave train is located in the first wave, varying with the granular material and the slope angle. For large slope angle, the amplitude of the first wave does not increase anymore and the remaining 
part of the energy is transferred in the following waves. An example is provided in figure \ref{fig:6} where the amplitude of the second wave increases with the slope angle 
whereas the amplitude of the first wave remains the same.

Our experiments illustrate that the total energy associated to the generated wave train does not 
depend on the granular media, but only on the parameter $\theta-\theta_c$. Understanding the exact dependance of the energy of each waves with the slope angle or the initial 
mass of granular material remains complicated. However, a first approach consists in evaluating the ratio of the energy of the generated wave 
train and the variation of the mechanical energy of the slide. Initially, the slide is at rest above the free surface, such that the total amount of energy transferred to 
the system is equivalent to the variation of its potential energy $\Delta E_m=E_p(t=0)-E_p(t_f)$, where:

\begin{eqnarray}
E_p(t=0) & = & m\,g\,\left[H+\frac{2}{3}\sqrt{\frac{2\,m\,\mathrm{tan}\theta}{\phi_i\rho_g\,L}}\right], \\
E_p(t_f) & = & \left(m - \phi_f\,\rho_w\,\mathcal{A}\,L\right)\,g\,z_G.
\end{eqnarray}

In these expressions, $m$ is the mass of the slide, $H$ denotes the water depth, $\theta$ is the slope, $L$ represents the width of the tank, $\mathcal{A}$ and $z_G$ are the 
area and the vertical position of the center of mass of the slide at the end of the collapse, respectively, $\phi_i$, $\phi_f$ are  the volumic fraction of the slide at the 
beginning and at the end of the collapse, respectively and 
$\rho_w$, $\rho_g$ are the density of water and the granular material, respectively. The volumic fractions of the slide at the 
beginning and at the end of the collapse, $\phi_i$ and $\phi_f$, can be expressed by :

\begin{eqnarray}
 \phi_i & = & \frac{m}{\rho_g\,\mathcal{A}(t=0)\,L} \\
 \phi_f & = & \frac{m}{\rho_g\,\mathcal{A}(t_f)\,L}
\end{eqnarray}

These expressions show that the ratio between the energy of the wave train and the total available energy in the system never exceeds 15\% regardless of the slope angle or the 
initial mass of the slide. The highest values occurred for high slope angle because the speed of the slide increases with the slope such that less energy is damped by the 
friction on the slope. Note, that these values are lower than those observed by Fritz et al. \cite{fritz2004} for impact at high Froude numbers, which confirms that taking into account the 
roughness of the slope drastically reduces the amount of energy transferred into the generated wave train. Understanding the quantity of energy lost by friction and how it is 
transferred in the several waves remains a challenging issue.


To conclude, we illustrate in this Letter the complex dynamic associated to the collapse of dry granular media into water with the objective of predicting the amplitude of the tsunami 
generated by a major cliff collapse. Our results uncovered some important findings such as the evolution of the wave during its propagation and the difference between dry and wet granular collapse. 
We have characterized the influence of the mass and we showed that the effect of the slope can be determined by rescaling the data with the variable $\theta-\theta_c$. 
However, the influence of water on dense flows of granular media remains an active problem and characterizing the interplay between the fluid and the slide during its collapse is a 
key issue to accurately predict the size of the generated waves. Examining the particular cases of collapse into deep water and collapse into a thin layer of water will be one of the aims of future research 
as these two extreme cases may simplify the situation. In addition, these experiments can be used as benchmark for future DEM-CFD simulations using a two-phase flow which would provide an insight into the dynamic of the granular beads \cite{topin2012}.

\section{Acknowledgements}
We thank Olivier Pouliquen, Pascale Aussillous and Pierre-Yves Lagr\'ee for fruitful discussions.

\bibliography{biblio}
\bibliographystyle{eplbib} 
\end{document}